\documentclass[12pt]{article}
\usepackage[utf8]{inputenc}
\usepackage{enumerate}
\usepackage{titling}
\usepackage{amsmath}
\usepackage{amsfonts}
\usepackage{bbm}
\usepackage{amsthm}
\usepackage{setspace}
\usepackage{subcaption}
 \RequirePackage{ifpdf}
\ifpdf
	\usepackage[pdftex]{graphicx}	
	\DeclareGraphicsExtensions{.pdf,.png,JPEG}
\else
	\usepackage{graphicx}		
	\DeclareGraphicsExtensions{.eps}
\fi
\graphicspath{ {./img/} } 

\usepackage{appendix}
\usepackage{authblk}
\usepackage{mathtools}
\usepackage{breqn}
\usepackage{mathrsfs}
\usepackage{xcolor}

\usepackage[left=2.5cm,right=2.5cm,top=2.5cm,bottom=2.5cm]{geometry}

\usepackage[sort, numbers]{natbib}

\usepackage{csquotes}

\newcommand{\R}{\mathbb{R}}

\newcommand{\ssq}{\subseteq}

\providecommand{\keywords}[1]
{
  \small	
  \textit{Key words:} #1
}

\title{Diffusion and Discrete Temporal Models of the Growth of Free-Ranging Cats in Urban Areas}
\author{Rodrigo Perusquía Cortés, Pablo Padilla Longoria}
\affil{IIMAS -- Universidad Nacional Autónoma de México}
\date{2024}

\begin{document}
\setstretch{1.2}

\maketitle
\begin{abstract}

The survival of the domestic cat (Felis catus) in various ecosystems has become increasingly relevant due to its impact on wildlife, public health, and society. In countries like Mexico, social factors such as abandonment have led to the feralization of the species and an unexpected increase in its population in urban areas. To design and implement effective population control methods, a thorough analysis of the species' population dynamics, along with the social factors influencing it, is necessary.


We propose a reaction-diffusion model to simulate the natural dispersal of the population within a bounded domain. After exploring the species' spreading ability, we construct a complex dynamical system based on the biological characteristics of cats and their intraspecific and interspecific interactions, which we explain and study in detail. Both deterministic and stochastic parameters are considered to enhance the realism of the simulations. Our results indicate that the population reaches equilibrium, highlighting the need for control methods combined with social regulations to achieve sustainability in the system.
\end{abstract}


\section{Introduction and Motivation}
\label{Intro}

The term "feral cat" refers to an unowned cat that commonly avoids or exhibits aggressive behavior toward people \cite{Dutcher} , while the term "free-roaming cat" refers to an unowned cat living outdoors \cite{Thompson}, both having limited interactions with humans. Feral cats form a subset of free-roaming cats, as the latter also includes abandoned, lost, and stray cats that were previously socialized to people  \cite{Flockhart}. Cats abandoned in the wild, apart from human socialization, along with the offspring of free-roaming cats, constitute the feral cat population \cite{Miller, Ogan}.

Since urban areas can have both free-roaming cats and owned cats allowed outdoors, we use the term "free-ranging cat" to refer to both categories. In this context, a free-ranging cat is defined as a cat that spends time outdoors with no physical limitations on its movement, regardless of its ownership status.


The growth of free-ranging cats in urban areas has become a significant problem in some countries due to loss of biodiversity, nuisance, transmission of diseases to other species, and the impairment of cats' welfare \cite{Gunther, Loss, Mella, Robertson}. In Mexico, for example, there is an estimated 28 million domestic animals (mainly dogs and cats), of which 70\% are homeless, with data indicating an annual increase of 20\%  \cite{VillagranAbandono2021}. However, this information has been repeatedly cited since 2016 \cite{FloresAbandono2016, GomezAbandono2018, MonrazAbandono2023, PatronAbandono2022}.


In 2021, INEGI (the Spanish acronym for the National Institute of Statistics and Geography) reported a total of 80 million owned pets, of which 43.8 million are dogs and 16.2 million are cats \cite{Enbiare}. However, there is no information available regarding an estimate of free-ranging domestic animals in Mexico, which would reflect the magnitude of the overpopulation problem of this species today.


Recent studies have been conducted to assess the impact of free-ranging dogs and cats on urban natural reserves in Mexico City, as well as the benefits of the control methods implemented \cite{Ramos}. Notably, cats have proven to be more resilient than dogs. This finding serves as additional motivation to study the population dynamics of free-ranging cats in depth and identify the reasons behind this phenomenon.


The first approach to modeling the population dynamics of a species and its growth is often through a system of ordinary differential equations (ODEs). First, one needs to identify the natural factors that govern population dynamics (e.g., birth, death, and migration rates) as well as the main factors that affect it (e.g., competition and social intervention) to include them in the model. To study the impact of the population in a physical domain, a more effective approach involves proposing a system of partial differential equations (PDEs) with boundary conditions.


An alternative approach to studying this topic is through a complex system based on graph theory and networks, where it is essential to establish the patterns and interactions between vertices with a high degree of precision to accurately reproduce the natural behavior of the species. In a complex system, we can distinguish between individuals, allowing us to study the population in a more personalized manner.


To conduct a comprehensive study of the growth of free-ranging cats, it is necessary to propose several approaches to this phenomenon. First, we will present a system of PDEs modeling the spatial growth of this population. This will complement the second model, which is a complex dynamical system that simulates the population dynamics of free-ranging cats, placing particular emphasis on its mating system and describing, with greater precision, the intrinsic factors that affect the population growth. A sensitivity analysis is performed on the second model to test its reliability, along with a brief study on the prediction of less well-known statistical population parameters. Finally, we draw conclusions from both models, discuss the most commonly implemented population control methods for free-ranging cats (such as lethal control, trap-neuter-return, and trap-vasectomy or hysterectomy-return), and establish connections between them.


\section{Reaction-Diffusion Model}
\label{RDM}

The system of ODEs formulated in \cite{Sharpe} provides an accurate description of the factors affecting the population of free-ranging cats in urban areas. Motivated by this, we also consider a subdivision of the population into kittens, adult females, and adult males for our model. Parameters such as birth rate $(b)$, sexual maturity $(m)$, intrinsic disease rates $(d_1, d_2, d_3)$, and intraspecific competition $(c_{2,2}, c_{2,3}, c_{3,2}, c_{3,3})$ are taken into account, along with factors related to abandonment $(a_1, a_2, a_3)$, animal rescue $(r_1, r_2, r_3)$, and other causes of death due to human intervention $(e_1, e_2, e_3)$, such as poisoning or vehicle accidents \cite{Miller}.

Let $\Omega \ssq \R^2$ be a bounded domain with a piece-wise smooth boundary endowed with unit outer normal \textbf{n} and $x_1(y,t), x_2(y,t)$ and $x_3(y,t)$ be the supopulation density of kittens, adult females and adult males, respectively. The reaction-diffusion system with Neumann boundary conditions
\begin{align}
\frac{\partial x_1}{\partial t}=& \frac{bx_2}{K_1 +x_2} + a_1 - x_1\bigg(p_1 +\frac{2m}{K_2+ x_1}\bigg), & y \in  \Omega, t > 0 \nonumber \nonumber \\[10pt]
\frac{\partial x_2}{\partial t}=& \frac{mx_1}{K_2 + x_1}  + a_2 -  x_2\big(p_2  + c_{2,2}x_2  +   c_{2,3}x_3 \big)  + D_2\Delta x_2, & y \in  \Omega, t > 0 \nonumber \\[10pt]
\frac{\partial x_3}{\partial t}=& \frac{mx_1}{K_2  + x_1}  +  a_3  - x_3\big(p_3  +   c_{3,2}x_2  +  c_{3,3}x_3 \big) +  D_3\Delta x_3, & y  \in  \Omega, t > 0 
\end{align}
\vspace{-20pt}
\begin{align}
\frac{\partial{x_1}}{\partial{\textbf{n}}}(y,t)& = 0, \hspace{15pt} \frac{\partial{x_2}}{\partial{\textbf{n}}}(y,t) = 0, \hspace{15pt} \frac{\partial{x_3}}{\partial{\textbf{n}}}(y,t)=0,  \hspace{30pt}& y   \in   \partial \Omega, t >  0 \nonumber \\[10pt]
x_1(y,0)& \geq 0, \hspace{15pt} x_2(y,0) \geq 0, \hspace{15pt} x_3(y,0) \geq 0,  \hspace{35pt}& y \in   \Omega, \nonumber
\end{align}
provides a model for the spatial growth of free ranging cats, where $D_2$ and $D_3$ are diffusion coefficients, $K_1$ and $K_2$ are half-saturation constants and $p_i := d_i + r_i + e_i$, for $i=1,2,3$, is the parameter of total decrease for the $i$-th subpopulation.


The proposed model includes Lotka-Volterra predator-prey interactions and Michaelis-Menten type growth. The Neumann boundary condition is based on the assumption that the region is closed or isolated, meaning that no individuals are entering or leaving it at the boundary. It is important to incorporate diffusion behavior into the system because domestic cats are natural predators and are considered an invasive species \cite{RodriguezRodriguez}. Additionally, system (1) assumes zero competition and diffusion coefficients in the first equation, as the impact of kittens on other subpopulations and their ability to move through the domain and survive are practically null until they reach adulthood or sexual maturity (approximately at 6 and 8 months, respectively) \cite{McCune}.

\begin{figure}[ht]
  \subcaptionbox*{{\footnotesize{a)}}}[.45\linewidth]{
    \includegraphics[width=\linewidth]{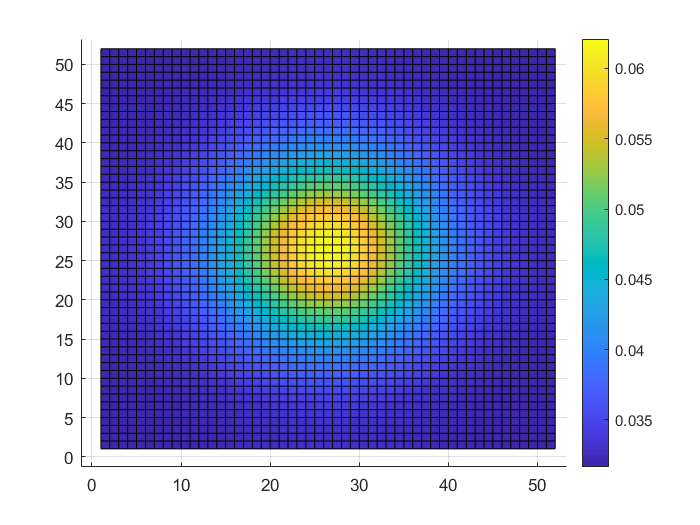}
  }
  \hfill
  \subcaptionbox*{{\footnotesize{b)}}}[.45\linewidth]{
    \includegraphics[width=\linewidth]{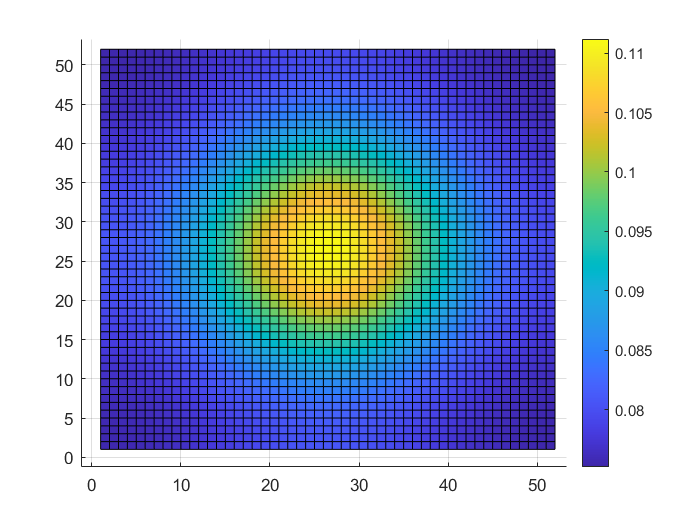}
  }
  \hfill
  \subcaptionbox*{{\footnotesize{c)}}}[.45\linewidth]{
    \includegraphics[width=\linewidth]{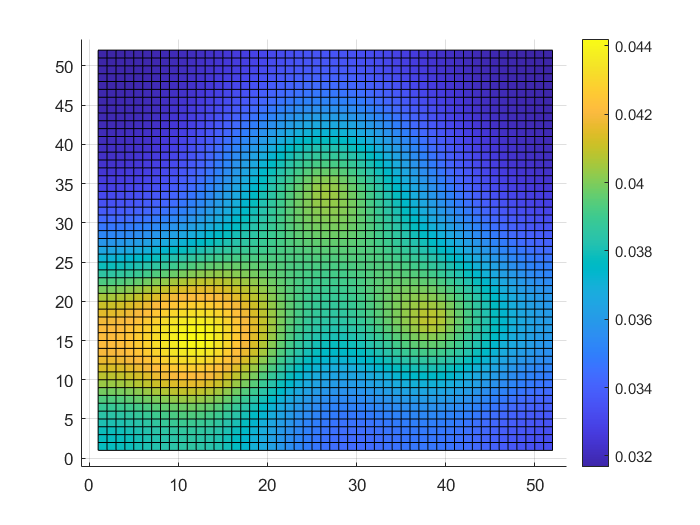}
  }
  \hfill
  \subcaptionbox*{{\footnotesize{d)}}}[.45\linewidth]{
    \includegraphics[width=\linewidth]{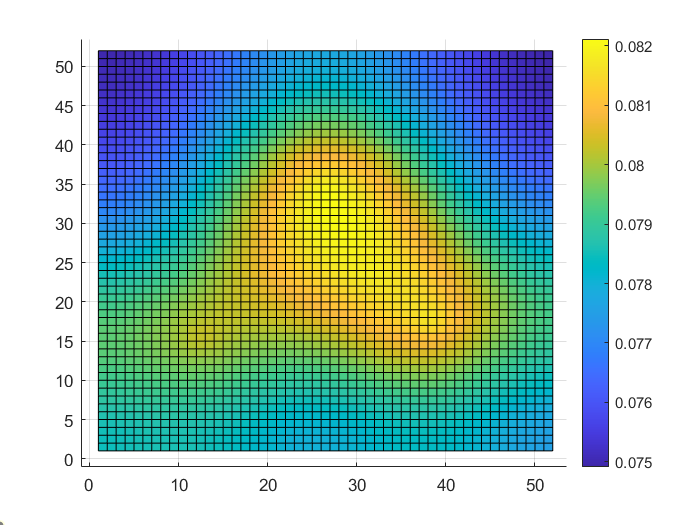}
  }

  \caption{
{\small{Illustrations of the spatial growth of free-ranging adult female cats (left figures) and adult male cats (right figures) when the species is introduced in a central subgrid (upper figures) or in several subgrids with different initial conditions (lower figures).}}}
  \label{Diff1}


\end{figure}

By implementing the five-point stencil discretization to approximate the Laplacian and the fourth-order Runge-Kutta method in time, while assuming that all parameters are constant in both time and space, we numerically simulate different biological situations in MATLAB R2024b using a  $50\times50$ grid mesh with the following fixed values: $b=2$, $m=0.5$, $a_1=\frac{3}{10}$, $a_2=\frac{1}{5}$, $a_3=\frac{1}{2}$, $p_1=0.9$, $p_2=0.5$, $p_3=0.5$, $c_{2,2}=0.5$, $c_{2,3}=0.3$, $c_{3,2}=0.7$, $c_{3,3}=0.9$, $K_1=1$, $K_2=0.5$, $D_2=2.5$, $D_3=4$ (Figure \ref{Diff1}).

This type of growth is characteristic of areas with abundant resources and no limitations on population spread. The inequality $D_3 > D_2$ affects the diffusion of subpopulations $x_2$ and $x_3$; specifically, male cats have a wider range for displacement in search of reproduction, while female cats tend to be more sedentary, as access to shelter and food resources for themselves and their kittens are their primary needs \cite{McCune}. Furthermore, the assumption $a_1, a_2, a_3 \neq 0$ contributes to the survival of the species, as both adult female and male cats reach a positive equilibrium.

In urban areas where settlements are divided by highways or avenues, which are surrounded by other natural areas with varying accessibility to resources, free-ranging cats exhibit certain spreading limitations depending on their location. Additionally, abandonment often occurs within the urban area. Therefore, it is interesting to re-simulate cases c) and d) from Figure 1, considering space-dependent diffusion and abandonment coefficients (see Figure 2).


Although the densities are more limited, different colonies tend to connect with each other to the extent that the surroundings allow. This behavior becomes particularly noteworthy when a colony is overpopulated and begins to access areas with abundant resources, especially in cities like Mexico City, which has over 20 nature reserves.

\begin{figure}[ht]
  \subcaptionbox*{{\footnotesize{a)}}}[.45\linewidth]{
    \includegraphics[width=\linewidth]{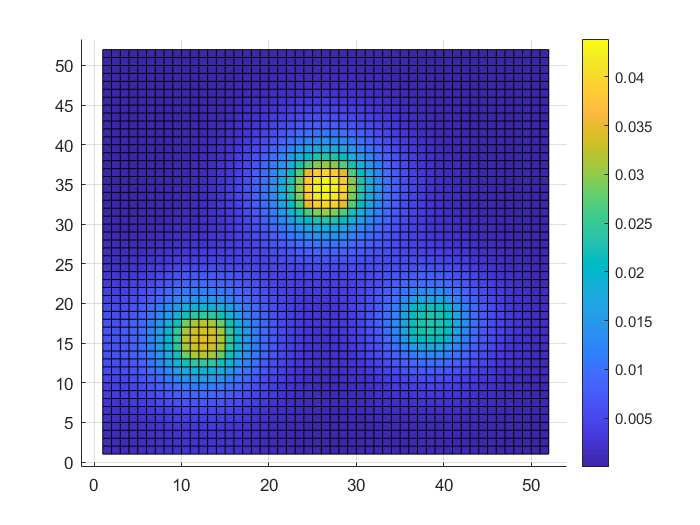}
  }
  \hfill
  \subcaptionbox*{{\footnotesize{b)}}}[.45\linewidth]{
    \includegraphics[width=\linewidth]{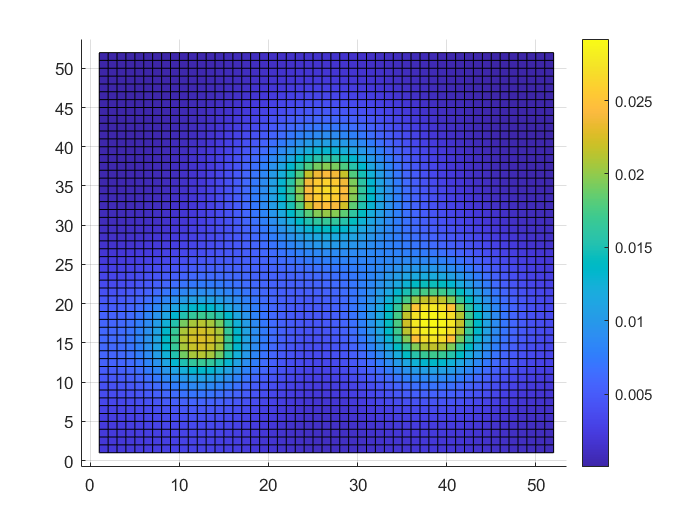}
  }

  \caption{
{\small{Spatial growth of free-ranging female cats (left figure) and male cats (right figure) is shown, considering different connected settlements in an urban area. We define $a_1, a_2, a_3, D_2, D_3$ as spatial step functions, where each colony and its surroundings have constant diffusion and abandonment parameters.}}}
\label{Diff2}
\end{figure}



The simulations of system (1) yield expected outcomes for the modeled phenomenon. An accurate example of the diffusive behavior of free-ranging cats is observed in the modeled density distribution of the species in Australia \cite{Legge}.


The existence of spatially homogeneous solutions for system (1) is possible due to the homogeneous Neumann boundary condition established. If the abandonment parameters are zero, except perhaps for $a_3$, there exists a homogeneous solution of the form 
for $(0,0,\hat{x}_3)$ for $\hat{x}_3 \in \R$; moreover, if $a_3=0$, then the origin is a solution of (1). It is concerning that the only way free-ranging cats might perish—meaning that the point of extinction constitutes an asymptotically stable equilibrium of the system—is when all the abandonment parameters are zero. However, this is a necessary condition but may not be sufficient.


The scenario of zero abandonment is hard to envision, as existing data demonstrates the magnitude of the situation in different countries. For example, data collected by PAOT (the Spanish acronym for the Environmental and Territorial Planning Attorney's Office) indicate that in Mexico, around 500,000 cats and dogs are abandoned each year \cite{VillagranAbandono2021}. As the results of the PDE model show, before attempting to implement a control method for the population, efforts should focus on reducing the abandonment rate. This can be achieved by enforcing the creation of laws against abandonment and ensuring their proper application.


\section{Complex System Model}
\label{CSM}

From a discrete perspective, we approximate the population dynamics of free-ranging cats in urban areas and study how this influences the growth and survival of the species in this environment. To achieve this, we construct a complex dynamical system in which vertices represent cats and edges represent their sexual interactions. In subsections \ref{MF} and \ref{SF}, we detail the most important features of free-ranging cats, along with the social factors that influence population dynamics. These factors are used to construct the complex dynamical model presented in subsection \ref{CSM}.


\subsection{Behavior and Reproduction}
\label{MF}

A free-ranging cat is generally a solitary and territorial animal that does not often trust people. They are solitary hunters, as their prey may not be sufficient to sustain a community; however, they can adapt to higher population densities if the environment provides ample resources \cite{McCune}. Free-ranging cats possess well-developed survival skills, such as the ability to fit into small spaces, climb trees or walls, and have an acute sense of hearing and smell, along with good low-light vision. These adaptations enable the species to thrive despite the challenges of their environment \cite{Dutcher, Little}. Additionally, their diet includes virtually anything they can catch, such as insects, birds, reptiles, fish, and smaller mammals, among others. \cite{Mella}.


Free-ranging cats compete with each other for territory. Male cats use a marking mechanism by spraying urine in their surroundings, and sometimes female cats in heat exhibit the same behavior. Gaining territory means access to more resources, shelter, and, particularly in the case of male cats, more female cats for reproductive purposes \cite{Burkholder, Griffin}.


The reproductive cycle of female cats is divided into proestrus, estrus, interestrus, diestrus, and anestrus. During proestrus, stage difficult to detect due to its short duration, they display affectionate behavior. Estrus follows proestrus and is the phase of receptivity to mating. Female cats are unique in that they are induced ovulators; both copulatory and non-copulatory stimulation, such as foreplay or environmental factors, increase the likelihood of ovulation during estrus \cite{Griffin}.


If ovulation does not occur, interestrus follows (if it is still the breeding season), marking the stage between consecutive estrus periods. Diestrus is the stage that occurs after a female cat has ovulated, in which physical and hormonal changes take place. When fertilization occurs, pregnancy follows diestrus. If fertilization does not happen, diestrus leads to pseudopregnancy, a condition that produces the same signs and symptoms as pregnancy but has a shorter duration \cite{Johnson, Little, McCune}.


Anestrus is a periodic phase of no cycling activity due to natural factors. The length of the breeding season varies by location; for example, in North America, reproductive activity occurs from February (peaking until April) to October or November, leading to winter anestrus due to the shorter daylight hours \cite{Little}.

Table 1 provides information about the average duration of the stages a female cat experiences in relation to its reproductive process. The values are drawn from registered data found in \cite{Griffin, Little, McCune, Stornelli, Veronesi}.
\begin{table}[h!]
  \begin{center}
   \caption{Female Cat Reproduction Data.}
    \label{FRD}
    \begin{tabular}{c|c}
      \textbf{Stage} & \textbf{Duration}\\
      \hline
      	Estrous Cycle & 14 days \\
	Estrous & 5-9 days \\
	Interestrous & 7 days\\
      	Gestation & 63 days \\
      	Lactation & 7 weeks \\
      	Weaning & 4 weeks \\
      	Pseudopregnancy & 45 days
    \end{tabular}
  \end{center}
\end{table}


Polygamy is the mating system of free-ranging cats \cite{Burkholder}. The onset of sexual maturity varies by gender (ranging from 8 to 18 months for male cats and from 4 to 18 months for female cats), breed, season, and physical conditions such as weight \cite{Little}. Even though estrous cycles become more irregular by the age of 8 years and sperm quality decreases as a male cat ages, cats can have offspring throughout their lifetime \cite{McCune}.


Female cats are seasonally polyestrous, which results in multiple pregnancies within a single reproductive season \cite{Griffin}. They can have from one to three litters per year, with a mean of 3.5 kittens per litter; furthermore, they can give birth to 50 to 150 kittens over a breeding life of 10 years \cite{Griffin}. According to \cite{Mccarthy}, the time to cycling after parturition is estimated to be 45 days. However, the time from the end of one gestation to the beginning of the next can vary depending on the individual; some female cats may experience a lactational anestrus that lasts up to 8 weeks after weaning, while others are capable of coming into estrus while still nursing \cite{Griffin, Little}.

\subsection{Social Factors}
\label{SF}

The lack of legislation that promotes animal welfare and the underenforcement of existing laws are well-documented issues \cite{Rodriguez}. Combined with social desensitization and ignorance on the topic, these factors constitute some of the main social contributions to the feralization of cats in urban areas.

As Satz notes,
\begin{displayquote}
``Throughout their lives, domestic animals rely on humans to provide them nourishment, shelter, and other care. The permanent dependency of domestic animals is created and controlled by humans, rendering them uniquely vulnerable to exploitation. Domestic nonhuman animals are, for this reason, perhaps the most vulnerable of all sentient beings." \cite{Satz} p. 80
\end{displayquote}


In Mexico, data collected by PAOT shows that around 500,000 cats and dogs are abandoned each year. According to \cite{Gabinete}, the main reasons attributed to the abandonment of domestic animals in Mexico are the lack of financial resources, responsibility, and time for care. Further studies revealed that cost issues were a leading factor in the abandonment of cats, and that four out of ten people who abandoned or relocated their cat did so because it required a significant commitment, while three out of ten cited a lack of space at home \cite{Mars}.

In 2021, Mexico City did not have any recorded criminal offenses related to the abandonment of domestic animals \cite{VillagranAbandono2021}. Furthermore, the COVID-19 pandemic worsened the situation, as indicated in \cite{MonrazAbandono2023}, leading to an increase in abandonment by at least 15\% during this period.


Free or low-cost neutering campaigns and rescue-adoption activities carried out by the government and other social organizations are some of the actions implemented in Mexico to combat the growth of free-ranging cats in cities and to raise awareness in society \cite{MonrazAbandono2023, PatronAbandono2022}. This is in addition to the recent creation and modification of laws promoting animal welfare \cite{Reforma}.

\subsection{A Complex System for the Growth of Free Ranging Cats}
\label{CSM}

We aim to describe the growth of free-ranging cats in an urban area by estimating the number of sexual interactions within the population through a discrete model that evolves over time. Its main components are the elements presented in sections \ref{MF} and \ref{SF}. We assume a weekly time step for the evolution, considering that the data presented in Table \ref{FRD} can be estimated in weeks.


Our complex dynamical system can be understood as a bipartite graph that changes its structure over time. Each vertex represents a cat and is initially labeled with its basic biological characteristics: gender and age. If the vertex represents a male cat (type I), a value for dominance is assigned; if it represents a female cat (type II), its reproductive cycle phase is recorded along with a pregnancy/pseudopregnancy counter (PC).


On one hand, female cats in heat attract male cats through vocalizations and their scent, often marking their territory by spraying urine \cite{Griffin, Little}. On the other hand, male cats have a wider home range in search of mating opportunities compared to female cats. Consequently, our graph is formed by directed edges from vertices of type I (male cats) to vertices of type II (female cats), simulating the natural sexual interactions between them. Since a dominant male is more likely to copulate than a less dominant one \cite{Mccarthy}, edges are created randomly with weights based on dominance and change as the system evolves, reflecting their polygamous mating system.

Let $t_0$ represent a week within the population's breeding season. At time $t_0$, the edges have the following considerations:
\begin{enumerate}
\item If a male cat (V1) encounters a female cat (V2), an edge (E1) is drawn from V1 to V2. This encounter may be for mating purposes or merely a casual interaction.
\item If another male cat (V3) encounters the same female cat (V2) and it has the appropriate conditions to become pregnant, an edge (E2) is drawn from V3 to V2. This strengthens the likelihood that (E1) represented a sexual encounter (it is also possible that V3 = V2).
\end{enumerate}


Edges are stored within the characteristic information of the adjacent vertices. Once the weekly interactions are completed, an analysis of the vertices of type II is conducted. If a female cat has the appropriate conditions to become pregnant and has one or more directed edges connecting to a vertex of type I that represents a male cat over the onset of sexual maturity, a process begins to determine whether it becomes pregnant, pseudopregnant, or neither. When one of the first two options occurs, the PC for the respective vertex starts running. If a female cat experiences pseudopregnancy, its PC records the duration of this process. If she achieves a successful pregnancy, her PC records the weeks of gestation, lactation, and weaning before resetting.


Vertices representing kittens should be considered once the weeks of gestation conclude. However, due to the high mortality rates during their early neonatal period (i.e., up to 6 or 8 weeks of age) \cite{Nutter}, they are added to the dynamic only when the PC of a pregnant female cat is reset. The addition of vertices due to social factors is also taken into account.


For the removal of vertices due to natural and social factors, a theoretical carrying capacity is established, as the species cannot grow indefinitely. Each vertex is assigned an individual survival probability $S$, and their product forms a joint survival probability $JS$. The selection of vertices follows two weighted random sampling procedures: the first depends on $JS$ and the second on their ages. The second procedure, referred to as $SS$, can be understood as the probability of being chosen among all vertices of the same type
Additionally, the number of vertices to be deleted is kept below an established upper bound that depends on the carrying capacity and the size of the population.


Afterwards, the time-dependent parameters are updated before all edges are removed from the dynamics as the system transitions from $t_0$ to $t_1$. If $t_1$ represents a week during the breeding season, the entire process described earlier is repeated. During the non-breeding season, only the addition and deletion of vertices, along with the update of time-dependent parameters, are carried out.

\subsection{Simulations}
\label{S}






We consider an initial condition of 50 free-ranging cats of equiprobable gender and randomly selected ages ranging from 7 to 36 months. We assume that our location is in North America, with a carrying capacity (CC) of 100 individuals, and that it is the start of the breeding season.


The parameters related to the population that we assume to be constant are as follows: onset of sexual maturity (8 months for both genders), duration of estrus and interestrus (one week each), and the durations for the breeding season, gestation, lactation, weaning, and pseudopregnancy (39, 9, 7, 4, and 7 weeks, respectively). Additionally, the age at which kittens are added to the dynamics is set at 10 weeks.


There is a scenario where one encounter leads to only one mating. When there is more than one encounter, we assume that there are more than two matings. Considering the ovulation and conception rates provided in \cite{Tsutsui}, which pertain to exactly one or three matings, we assume a constant rate of induced ovulation and conception: for one encounter, the rates are 72\% and 37\%, respectively; for more than one encounter, the rates are 85\% and 93\%, respectively.


A combination of constant and stochastic values is considered for social factors. We assume a weekly addition of one vertex following a Bernoulli distribution $Ber\big(p=\frac{1}{2}\big)$. This addition may represent an abandoned or missing cat, or an owned cat with access to the outdoors. The most common immediate scenarios for this individual are survival and adaptation, death or displacement, rescue or recovery, and return to its home on its own. Thus, the immediate continuity of the vertex in the dynamics follows a Bernoulli distribution $Ber\big(p=\frac{1}{4}\big)$.


Deterministic parameters are also taken into account in the model. Dominance is a complex value to define, as it is related to age, physical condition (including body weight and size), and breed, among other factors. Since male cats attain peak reproductive function at 2 years old \cite{Burkholder} and we assume that all male cats reach sexual maturity at $8$ months, we define dominance as a function of age $A$ measured in months: $D = \max\{ \min\big(1, \frac{A - 8}{16}\big), 0\}.$ A similar estimation is used in \cite{Mccarthy}.


Let $TP$ represent the total population, $AC$ the number of adult cats (age $\geq 6$ months) and $YC$ the number of young cats in the respective week. The daily survival probability assigned to each vertex is given by $S := .9991 - \frac{(.9991 - .99804)AC + (.9991-.99233)YC}{CC}$, which is a slight variation of the one considered in \cite{Miller}. The weekly joint survival probability is defined as $JS = S^{7TP}$, and the proposed upper bound for the weekly removal of vertices is $U = \max(1, \left \lfloor{\frac{2TP}{CC}}\right \rfloor)$.


Stochasticity plays an important role in the model, defining parameters that either have support from well-known data or do not. For example, litter size is determined through a weighted random sampling procedure: the sizes considered range from 3 to 6, with weights chosen to ensure that the expected value is 3.5, aligning with real data \cite{Griffin, McCune}. The removal of kittens follows a similar procedure: for each litter of size $L$, the number of kittens to be removed is selected from 0 to $L$, with weights designed to achieve an expected value of $0.45L$. This is based on the fact that approximately 55\% of kittens survive after ten weeks \cite{Nutter}.


We assume that the removal weighing procedure $SS$, mentioned in section \ref{CSM}, follows a gamma distribution $\Gamma\big(5, \frac{1}{2}\big)$. This distribution is light-tailed, making it more likely for the youngest and oldest free-ranging cats to be removed from the dynamics compared to those around two years old (the peak of their reproductive function). Additionally, we consider the immediate removal of individuals older than six years, as it is uncommon to see them in urban areas \cite{Ogan}.



Since our goal is to replicate the reproduction of free-ranging cats, the edges in the dynamics represent sexual interactions, which are drawn only during the breeding season. We categorize these interactions into primary (edges of type I) and secondary (edges of type II).
\begin{figure}[ht]
  \subcaptionbox*{{\footnotesize{a) Edges of type I}}}[.45\linewidth]{
    \includegraphics[width=\linewidth]{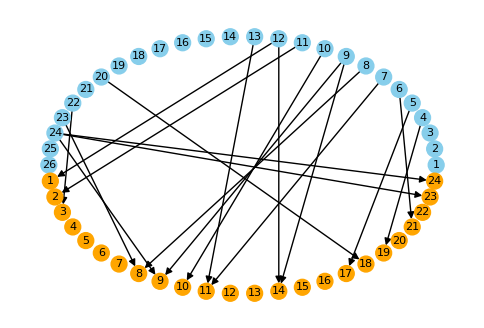}
  }
  \hfill
  \subcaptionbox*{{\footnotesize{b) Edges of type II}}}[.45\linewidth]{
    \includegraphics[width=\linewidth]{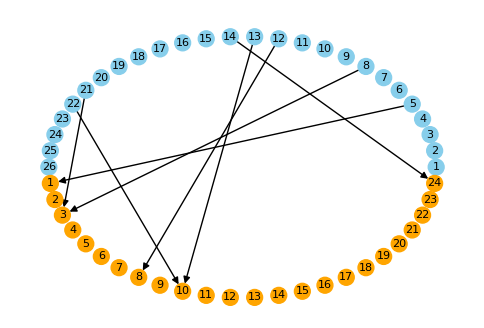}
  }
  \caption{{\small{Division of edges in a one-week simulation between female cats (orange vertices) and male cats (cyan vertices).}}}
  \label{InteractionsDivision}
\end{figure}

\newpage
Let $EP$ represent the number of female cats in estrus and $DD$ the number of highly dominant male cats. We define a male cat as highly dominant when its dominance $D$ satisfies $D \geq 0.75$. It is reasonable to expect that at least once a week, all female cats in estrus or all highly dominant males engage in a sexual interaction. So, the number of weekly primary interactions $E$ is randomly selected between $\min(EP, DD)$ and $\max(EP, DD)$. 


To determine the adjacent vertices for an edge of type I, a vertex of type I is selected with a weight equal to its dominance, and a vertex of type II is chosen with a weight based on its reproductive phase, as highly dominant male cats and female cats in estrus are more likely to encounter each other. Additionally, to prevent this stochasticity from resulting in multiple edges being drawn from the same vertex of type I to too many vertices of type II, or vice versa, we impose an upper bound on the number of edges attached to each vertex.


We say that a vertex of type II is eligible if it represents a female cat in estrus and is adjacent to a vertex of type I. Let $C$ be the total number of eligible vertices of type II. Since it is possible that not all eligible female cats engage in secondary interactions, but a fraction of them does, we assume that the number of edges of type II is a random value bounded by $\frac{C}{3}$ and $\frac{C}{2}$.


The drawing of edges of type II proceeds as follows: a vertex of type I with an age greater than 8 months and an eligible vertex of type II are randomly chosen. Secondary interactions reaffirm that a primary interaction was sexual in nature and demonstrate that free-ranging cats are polygamous. Moreover, these interactions allow male cats with lower dominance to participate in the dynamics.




\vspace{-10pt}
\begin{figure}[ht]
  \subcaptionbox*{{\footnotesize{a) $1^{st}$ week}}}[.43\linewidth]{
    \includegraphics[width=\linewidth]{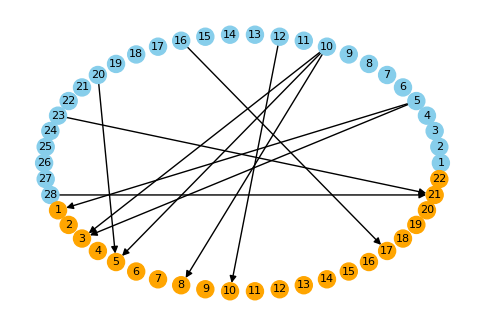}
  }
  \hfill
  \subcaptionbox*{{\footnotesize{b) $14^{th}$ week}}}[.43\linewidth]{
    \includegraphics[width=\linewidth]{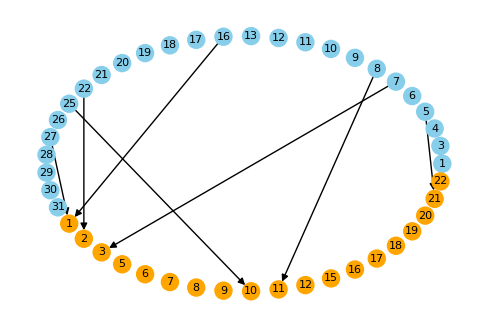}
  }
  \subcaptionbox*{{\footnotesize{c) $27^{th}$ week}}}[.43\linewidth]{
    \includegraphics[width=\linewidth]{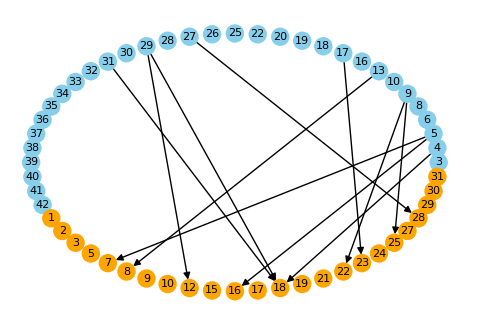}
  }
  \hfill
  \subcaptionbox*{{\footnotesize{d) $40^{th}$ week}}}[.43\linewidth]{
    \includegraphics[width=\linewidth]{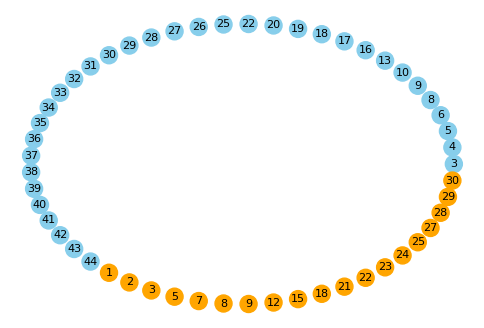}
  }
  \caption{{\small{Evolution of edges throughout the breeding season, without any divisions.}}
}
  \label{InteractionsEvolution}
\end{figure}

\begin{figure}[ht]
  \subcaptionbox*{{\footnotesize{a) $1^{st}$ week}}}[.45\linewidth]{
    \includegraphics[width=\linewidth]{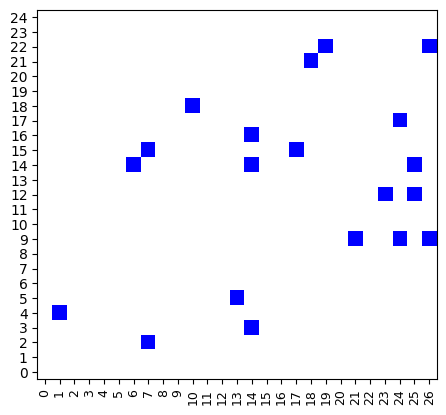}
  }
  \hfill
  \subcaptionbox*{{\footnotesize{b) $32^{th}$ week}}}[.45\linewidth]{
    \includegraphics[width=\linewidth]{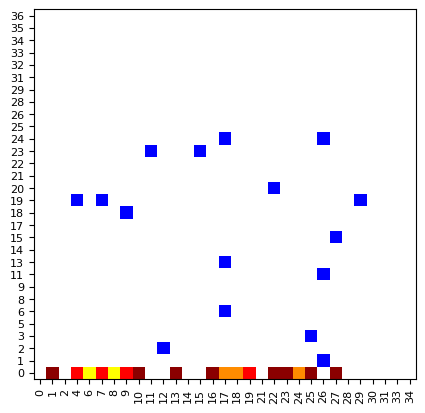}
  }
  \caption{{\small{The PC color bar consists of a total of five colors. A white tile is associated with a female cat that has not yet ovulated. When ovulation occurs, two sequences of colors may appear in the dynamics: White $\stackrel{{\footnotesize\text{1 week}}}{\longrightarrow}$ Dark Red $\stackrel{{\footnotesize\text{7 weeks}}}{\longrightarrow}$ Red $\stackrel{{\footnotesize\text{2 weeks}}}{\longrightarrow}$ Dark Orange $\stackrel{{\footnotesize\text{7 weeks}}}{\longrightarrow}$ Yellow $\stackrel{{\footnotesize\text{4 weeks}}}{\longrightarrow}$ White, for pregnant cats, or White $\stackrel{{\footnotesize\text{1 week}}}{\longrightarrow}$ Dark Red $\stackrel{{\footnotesize\text{7 weeks}}}{\longrightarrow}$ White, for pseudopregnant cats.}}}
  \label{Tiles}
\end{figure}

Another way to illustrate the complex dynamical system is through a tile representation, where edges between vertices $i$ (type I) and $j$ (type II) are represented by colored tiles (blue) with coordinates $(i,j)$. Additionally, a color bar is included in this representation to symbolize the PC of each female cat (see Figure \ref{Tiles}).

In the tile representation, we can identify biological phenomena within the population. For example, the introduction of kittens influences the dynamical behavior of the population. In Figure \ref{Tiles} b), the PC color bar ends with many white tiles because female kittens cannot become pregnant until they reach sexual maturity. Additionally, blue tiles are concentrated in the lower left side of the illustration, indicating that adult cats tend to monopolize the interactions.

\section{Results and Sensitivity Analysis}
\label{RSS}

\subsection{Population Behavior}
\label{RR}


The proposed complex dynamical model allows us to study the growth of the population over time. Since the model considers stochastic parameters, it produces different results with each execution of the code. Therefore, it is important to implement the model a sufficiently large number of times to obtain expected outcomes and draw meaningful conclusions. We conduct 1,000 simulations over a 10-year period, taking the average weekly total population from all the tests.


Since the simulations start in the first week of the breeding season, it takes some time for the population to increase. Afterward, from approximately the second year onward, a periodic growth becomes noticeable, which is typical behavior for seasonal breeders (see Figure 6).


The carrying capacity (CC) is reached for the first time around the third year, and this occurs periodically in subsequent years. After reaching CC, the population continues to grow but at a slower rate. When the population peaks, it begins to decrease almost automatically.


The value $U$ established as an upper bound for the weekly removal of vertices, is proposed as a function of half the carrying capacity. This is motivated by the fact that the logistic growth model for populations experiences a change in the growth rate when this value is reached. This, combined with the inverse relationship between the weekly joint survival probability and the population size, facilitates effective self-regulation of the population around the carrying capacity.

\begin{figure}[!ht]
\begin{center}
\includegraphics[width=.7\textwidth,keepaspectratio]{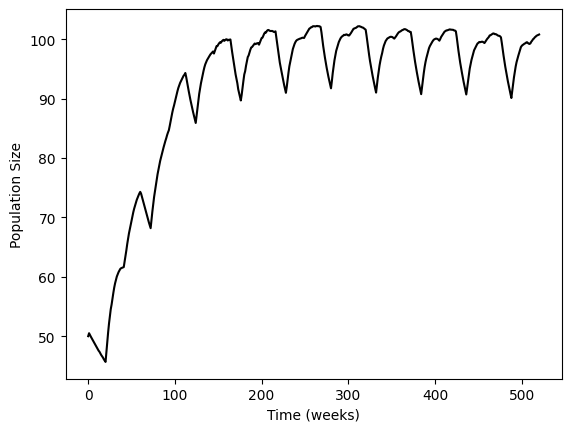}
\vspace{-10pt}
\caption{{\small{The population of free ranging cats follows a periodical growth behavior.}}}
\label{1000}
\end{center}
\end{figure}
\vspace{-10pt}

\subsection{Sensitivity Analysis on Interactions}
\label{BS}

Many parameters of the complex dynamical system are based on real data found in existing literature, supporting the model's reliability. However, there are values for which we had to make assumptions that, at first glance, might yield unrealistic results. Probably the most significant of these is the estimate for the number of interactions.


If our assumptions overestimate the weekly number of primary sexual interactions between male and female cats, significant changes in the outcomes should occur if these interactions decrease. To understand the extent of the impact of our estimates on the total number of individuals, we perform a sensitivity study. We consider the value $\frac{E}{n}$, for $n=1, 2, ..., 10$, as the number of primary interactions. For each $n$ we conduct 100 simulations over a 10-year period.


\begin{figure}[ht]
  \subcaptionbox*{{\footnotesize{a)}}}[.45\linewidth]{
    \includegraphics[width=\linewidth]{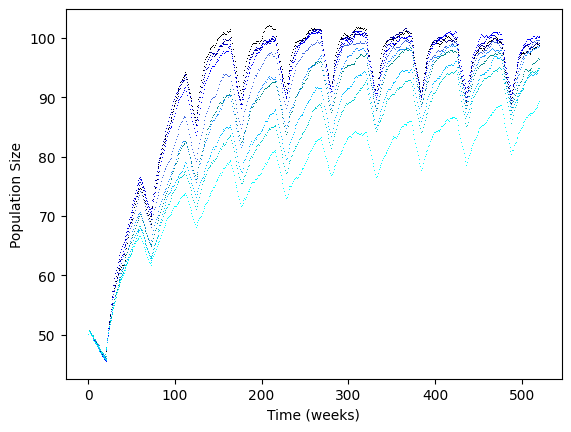}
  }
  \hfill
  \subcaptionbox*{{\footnotesize{b)}}}[.45\linewidth]{
    \includegraphics[width=\linewidth]{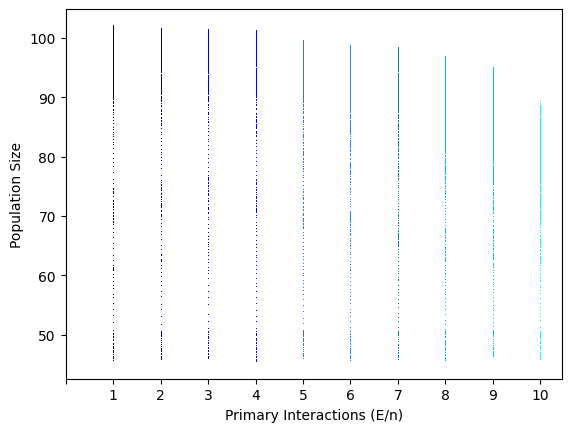}
  }
\vspace{-5pt}
  \caption{{\small{The graphs in a) correspond to the dotted segments in b) by tone; the lighter the tone, the greater the reduction in encounters considered for the simulations.}}}
  \label{Stability}
\end{figure}

\vspace{-5pt}

Regardless of the reduction in interactions, a periodic growth behavior is observed in each case (see Figure \ref{Stability} a)). As expected, fewer interactions directly impact the growth of the population. However, it is noteworthy that this impact remains significant until the primary encounters are reduced to one-fifth of their original number, at which point the population does not quite reach the carrying capacity, although it comes close (see Figure \ref{Stability} b)).

\begin{figure}[!ht]
\begin{center}
\includegraphics[width=.60\textwidth,keepaspectratio]{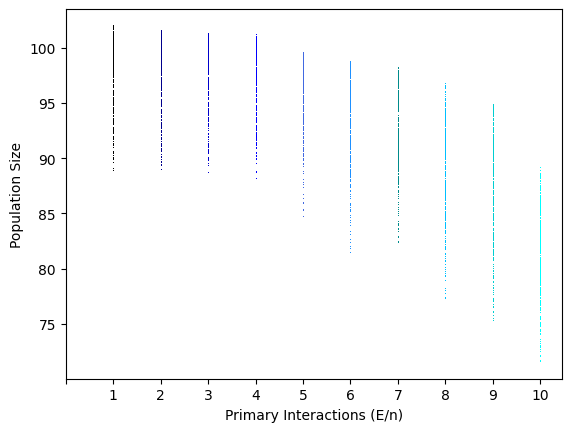}
\vspace{-8pt}
\caption{{\small{This illustration can be viewed as an amplification and refinement of Figure \ref{Stability} b), as it records the population size from the third year onward.}}}
\label{S0}
\end{center}
\end{figure}

After the third year in the simulations, the periodic growth of the population begins to stabilize. In contrasting the cases of no reduction and a reduction to one-tenth of the interactions, there is a significant variation in the total population, with a difference of nearly 20 individuals between the respective minimum and maximum values reached (see Figure \ref{S0}).

\begin{figure}[ht]
  \subcaptionbox*{{\footnotesize{a) Population growth behavior}}}[.45\linewidth]{
    \includegraphics[width=\linewidth]{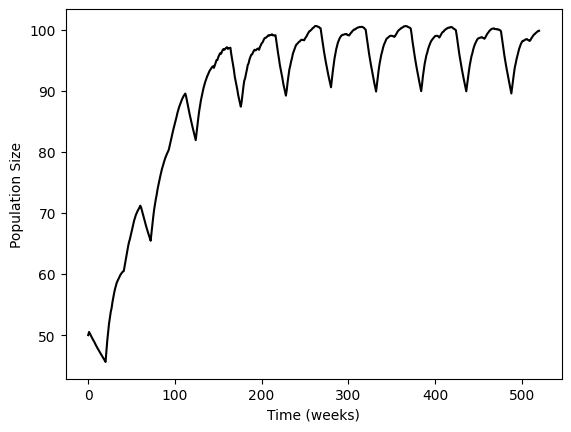}
  }
  \hfill
  \subcaptionbox*{{\footnotesize{b) Sensitivity Analysis}}}[.45\linewidth]{
    \includegraphics[width=\linewidth]{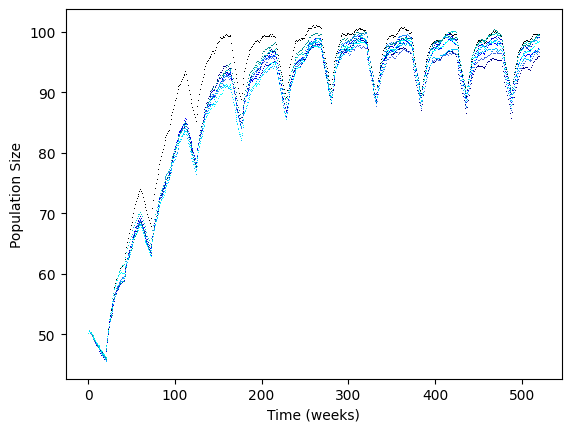}
  }
  \caption{{\small{Results for the variation of secondary interactions are presented. In panel a), there is no significant change compared to the results in Figure \ref{1000}, except for an initially smaller growth rate. In panel b), we observe similar outcomes regardless of the reduction rate.}}}
  \label{Stability2}
\end{figure}

For secondary interactions, we study the impact of removing the lower bound of our estimate by performing 1,000 simulations over a 10-year period (Figure \ref{Stability2} a)). Additionally, we conduct a similar sensitivity analysis by reducing the secondary interactions while keeping the original estimate of the number of primary interactions fixed (Figure \ref{Stability2} b)).

Results show that the number of sexual interactions between male and female cats is an important factor contributing to the maximum population increase. Nevertheless, it is not as significant as the intrinsic biological features of the species, such as the short durations of gestation, lactation, and weaning, as well as the high ovulation and conception rates of female cats.




A reduction in interactions between female and male free-ranging cats can naturally serve as a control method for the population; however, a substantial decrease is necessary to achieve significant results. Unfortunately, preventing cats from interacting is nearly impossible, especially when a large number are already established in an area.

\subsection{Statistical Parameters}
\label{FC}

The life expectancy of a free-ranging cat is not well understood due to various influencing factors, including diseases, ownership status, urban landscape, and access to resources \cite{Dutcher}. According to \cite{Miller}, it is estimated to be less than five years, while \cite{Loyd} indicates that it is around three years. Other demographic parameters, such as the mean, median, and mode age of the population, are also challenging to determine.

\begin{figure}[!ht]
\begin{center}
\includegraphics[width=.57\textwidth,keepaspectratio]{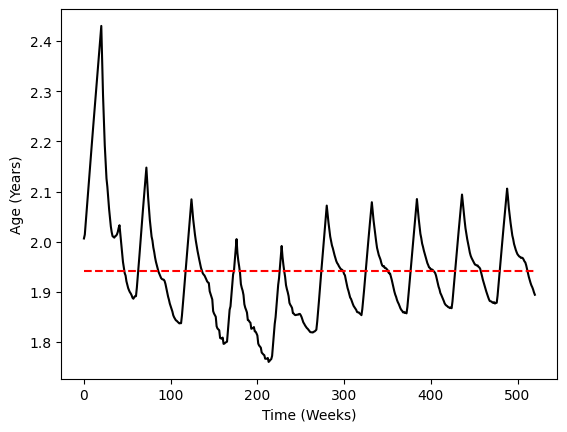}
\vspace{-5pt}
\caption{{\small{Weekly mean age (black continuous line) and overall mean age (red dashed line).}}}
\label{MA1}
\end{center}
\end{figure}

According to our model, after performing 1,000 simulations and taking the average values of all outcomes, the weekly mean age of the population exhibits a periodic behavior attributed to the seasonal breeding of free-ranging cats. The mean age decreases when litters are produced and increases during the rest of the time. Stability is reached after approximately five years, with the total mean age nearing two years (see Figure \ref{MA1}).

\begin{figure}[ht]
  \subcaptionbox*{{\footnotesize {a) Ages of individuals removed weekly (for one simulation)}}}[.45\linewidth]{
    \includegraphics[width=\linewidth]{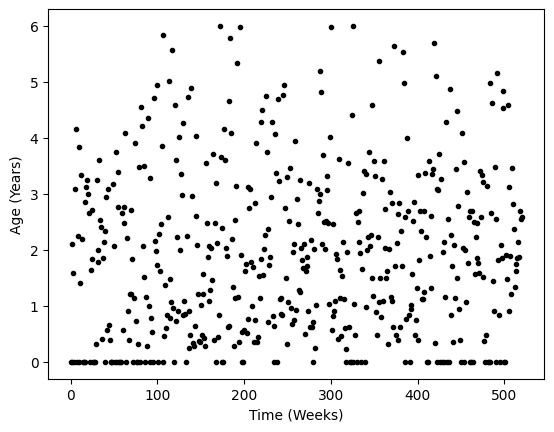}
  }
  \hfill
  \subcaptionbox*{{\footnotesize {b) Weekly mean age (black continuous line) and overall mean age at death (red dashed line).}}}[.45\linewidth]{
    \includegraphics[width=\linewidth]{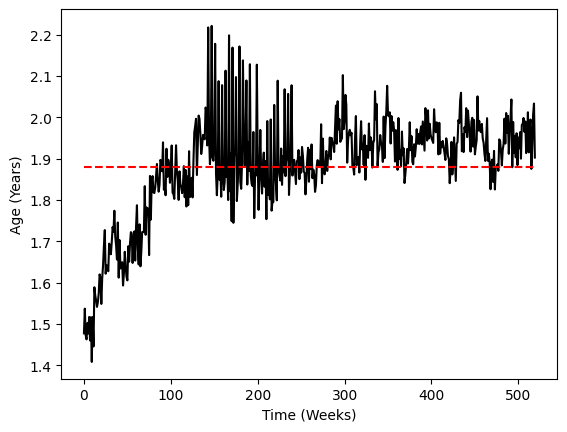}
  }
\vspace{-5pt}
  \caption{{\small{In panel a), individuals of various ages were removed from the complex system. Dots at zero height indicate that no vertices were removed during the respective week. In panel b), the chaotic behavior is attributed to the age variability at which individuals die across 1000 simulations.}}}
  \label{Mort}
\end{figure}

During each simulation, the ages of the individuals removed weekly were recorded (see Figure \ref{Mort} a)). We attribute the concentration of data in the lower part of the illustration to the low survival rate of young individuals and the difficulty free-ranging cats face in surviving many years. The weekly and overall mean ages at death across all simulations can be seen in Figure \ref{Mort} b). Although the weekly mean age at death is challenging to predict, it tends to stabilize over time. Notably, the proximity of the mean age, the mean age at death, and the mode of $SS$ is an important result.

\newpage
\section{Discussion}
\label{D}

The three most studied and implemented control methods for the population of free-ranging cats are lethal control (LC), trap-neuter-return (TNR), and trap-vasectomy/hysterectomy-return (TVHR). Each method leads to a reduction in the population's interactions, which is shown to be a natural control method in section \ref{BS}. LC reduces the number of interactions through the permanent removal of individuals; TNR suppresses sexual behaviors; and TVHR causes non-effective sexual encounters, as the sexual instinct of cats is not altered at all.


LC has been shown to be an ineffective population control method unless there is a high capture rate \cite{Mccarthy}. The most significant limitation of LC is that if just two individuals encounter and mate, the population can continue reproducing. Additionally, this method is controversial because it involves taking lives. TNR and TVHR are more widely accepted in society, and each method reduces the probability of reproduction due to the presence of both intact and altered individuals in the population.


The analysis in \cite{Miller} shows that TVHR is more effective than TNR. Although TNR and TVHR are similar control methods, they lead to different dynamics. In TNR, intact male cats redirect to intact female cats, as neutered female cats are part of the population. In contrast, with TVHR, intact male cats cannot distinguish between intact and altered female cats, resulting in a positive rate of non-effective sexual encounters, which contributes to the decline of the population.


Although both TNR and TVHR have been proven effective as population control methods in practice \cite{Levy, Mendes}, they also have disadvantages. In the case of female cats, hormonal activity continues after hysterectomy, and long-term exposure to sexual hormones is a risk factor for the development of mammary cancer \cite{Munson}, while ovariohysterectomy significantly decreases this risk \cite{Overley}; the earlier it is performed, the better. Nevertheless, TNR increases the life expectancy of cats, resulting in prolonged exposure of wildlife to cats, which are known to be surplus killers \cite{Loyd2, Mccarthy2}.


Reaching sustainability in a system does not necessarily mean eradicating the population of free-ranging cats; for example, they can contribute to controlling various pests (e.g., rodents, some insects, pigeons). Since there is no perfect control method, we emphasize the necessity for innovative regulatory policies. These could involve a strategic combination of the most common control methods, leveraging the dynamics of the free-ranging cat population.

\section{Conclusions}
\label{C}

The presented models address fundamental questions about the population of free-ranging cats, such as how it grows in space and why it is self-sustaining. Both models are based on biological and social features related to the population, incorporating registered data, and the results obtained enhance the reliability of each model.


On one hand, the PDE model effectively addresses the displacement abilities of free-ranging cats and how different colonies may connect through the population's diffusive component. On the other hand, the complex system more accurately captures the periodic growth behavior of the population. Future works should consider modifications to the time-dependent (periodic) parameters in the PDE system to achieve stability in a periodic solution, ensuring that the growth behavior is accurately represented. Furthermore, additional implementations in the complex model should explore dividing individuals into different colonies and incorporating migration parameters between them, as discussed in \cite{Miller}.


Other studies related to the dynamics of free-ranging cats have used daily or six-month time steps \cite{Miller, Mccarthy, Miller2}. To our knowledge, this is the first time a weekly time step has been proposed for our complex system model, along with the prediction of statistical parameters for the population—data that has not been extensively explored. Further analyses should be conducted within the model, considering alternative distributions for the SS parameter, such as exponential, Weibull, or log-normal. Different distributions may be better suited to various real-world scenarios, and predicting additional measures of central tendency would provide valuable complementary insights.



The continuous model highlights the need for social regulation in response to the issue of abandonment. Creating laws and ensuring their proper enforcement is essential, and additional measures, such as low-cost sterilization campaigns and the rescue and adoption of free-ranging cats, should also be considered. Furthermore, the analysis of the discrete model reveals that intrinsic factors of the population (e.g., high ovulation and conception rates) enable it to self-sustain around the carrying capacity of the inhabited area. Therefore, it is crucial to design and implement effective population control methods.


The implementation of a control method in both models is an important area to explore in detail. The PDE model can be studied through the theory of Optimal Control of PDEs, where control variables are introduced into the system along with a functional to optimize, typically representing the costs associated with the control policy. A similar study is conducted in \cite{Perus} using the theory of Optimal Control of ODEs. In the case of the complex system, the proposed control method should be integrated into the dynamics at varying rates to predict its effectiveness. Comparable studies are carried out in \cite{Miller, Mccarthy}, which compare different regulation methods.

\vspace{30pt}

\textbf{Conflict of interest}

{The authors declare no conflict of interest.}

\newpage


\addcontentsline{toc}{chapter}{Referencias}
\bibliographystyle{apalike}
\bibliography{Articulo}

\begin{thebibliography}{}

\bibitem[Gab, 2014]{Gabinete}
 (2014).
\newblock Maltrato y abandono de animales.
\newblock Gabinete de Comunicación Estratégica.
\newblock
  https://gabinete.mx/images/reportes/2014/sociedad/info\_maltrato\_abandono\_animales\_2014.pdf.

\bibitem[Ref, 2021]{Reforma}
 (2021).
\newblock Dictamen de la comisión de justicia de diversas iniciativas con
  proyecto de decreto que reforman y adicionan diversas disposiciones del
  código penal federal, en materia de sanción del matrato animal.
\newblock Cámara de Diputados LXIV Legislatura. Comisión de Justicia.
\newblock
  https://www.diputados.gob.mx/LeyesBiblio/iniclave/64/CD-LXIV-III-2P-354/02\_dictamen\_354\_23mar21.pdf.

\bibitem[Enb, 2021]{Enbiare}
 (2021).
\newblock Presenta inegi resultados de la primera encuesta nacional de
  bienestar autorreportado (enbiare) 2021.
\newblock Comunicado de prensa núm 772/21.
\newblock
  https://www.inegi.org.mx/contenidos/saladeprensa/boletines/2021/EstSociodemo/ENBIARE\_
  2021. pdf.

\bibitem[Mar, 2022]{Mars}
 (2022).
\newblock Resultados del estudio actitudinal - méxico. Índice de las mascotas
  sin hogar.
\newblock Mars Petcare.
\newblock
  https://mex.mars.com/sites/g/files/jydpyr496/files/2022-10/Indice\_de\_Mascotas\_sin\_Hogar\_Actitudinal.pdf.

\bibitem[Burkholder et~al., 2015]{Burkholder}
Burkholder, T., Ledesma-Feliciano, C., Vandewoude, S., and Baker, H. (2015).
\newblock {\em Biology and Diseases of Cats}, pages 555--576.

\bibitem[Dutcher et~al., 2022]{Dutcher}
Dutcher, A., Pias, K., Sizemore, G.~C., and Vantassel, S.~M. (2022).
\newblock Free-ranging and feral cats.
\newblock Wildlife Damage Management Technical Series. USDA, APHIS, WS National
  Wildlife Research Center. Fort Collins, Colorado.
\newblock 25 pp.

\bibitem[Flockhart and Coe, 2018]{Flockhart}
Flockhart, D. T.~T. and Coe, J.~B. (2018).
\newblock Multistate matrix population model to assess the contributions and
  impacts on population abundance of domestic cats in urban areas including
  owned cats, unowned cats, and cats in shelters.
\newblock {\em PLOS ONE}, 13(2):1--34.

\bibitem[Flores~Escalera et~al., 2016]{FloresAbandono2016}
Flores~Escalera, H.~E., Merodio~Reza, L.~G., Gastélum~Bajo, D.~H.,
  Díaz~Salazar, M.~C., Ríos de~la Mora, I.~S., Domínguez~Arvizú, M.~H.,
  Acosta~Islas, A., de~la Torre~Valdez, Y., and Ayala~Ríos, E. (2016).
\newblock Proposiciones con punto de acuerdo que exhorta a los gobiernos
  estatales a realizar campañas de esterilización de perros y fomenten
  campañas de adopción de perros en situación de calle.
\newblock Sistema de Información Legislativa de la Secretaría de
  Gobernación.

\bibitem[Griffin, 2001]{Griffin}
Griffin, B. (2001).
\newblock Prolific cats: The estrous cycle.
\newblock {\em Compendium on Continuing Education for the Practicing
  Veterinarian}, 23:1049--1056.

\bibitem[Gunther et~al., 2015]{Gunther}
Gunther, I., Raz, T., Berke, O., and Klement, E. (2015).
\newblock Nuisances and welfare of free-roaming cats in urban settings and
  their association with cat reproduction.
\newblock {\em Preventive Veterinary Medicine}, 119(3):203--210.

\bibitem[Gómez~Álvarez, 2018]{GomezAbandono2018}
Gómez~Álvarez, D. (2018).
\newblock Proposición con punto de acuerdo por el que el senado de república,
  solicita respetuosamente a la cámara de diputados considerar una ampliación
  al presupuesto asignado a la secretaria de salud, en específico para el
  centro nacional de programas preventivos y control de enfermedades
  (cenaprece) en materia de esterilización canina y felina, así como de
  vacunación antirrábica. además de incrementar el fondo de aportaciones
  para los servicios de salud (fassa), con el objetivo fortaler los programas
  de zoonosis en las entidades federativas.
\newblock Gaceta del Senado.

\bibitem[Ireland and Miller~Neilan, 2016]{Miller}
Ireland, T. and Miller~Neilan, R. (2016).
\newblock A spatial agent-based model of feral cats and analysis of population
  and nuisance controls.
\newblock {\em Ecological Modelling}, 337:123--136.

\bibitem[Johnson, 2022]{Johnson}
Johnson, A.~K. (2022).
\newblock Normal feline reproduction: The queen.
\newblock {\em Journal of Feline Medicine and Surgery}, 24(3):204--211.
\newblock PMID: 35209768.

\bibitem[Legge et~al., 2017]{Legge}
Legge, S., Murphy, B., McGregor, H., Woinarski, J., Augusteyn, J., Ballard, G.,
  Baseler, M., Buckmaster, T., Dickman, C., Doherty, T., Edwards, G., Eyre, T.,
  Fancourt, B., Ferguson, D., Forsyth, D., Geary, W., Gentle, M., Gillespie,
  G., Greenwood, L., Hohnen, R., Hume, S., Johnson, C., Maxwell, M., McDonald,
  P., Morris, K., Moseby, K., Newsome, T., Nimmo, D., Paltridge, R., Ramsey,
  D., Read, J., Rendall, A., Rich, M., Ritchie, E., Rowland, J., Short, J.,
  Stokeld, D., Sutherland, D., Wayne, A., Woodford, L., and Zewe, F. (2017).
\newblock Enumerating a continental-scale threat: How many feral cats are in
  australia?
\newblock {\em Biological Conservation}, 206:293--303.

\bibitem[Levy et~al., 2003]{Levy}
Levy, J., Gale, D., and Gale, L. (2003).
\newblock Evaluation of the effect of a long-term trap-neuter-return and
  adoption program on a free-roaming cat population.
\newblock {\em Journal of the American Veterinary Medical Association},
  222:42--6.

\bibitem[Little, 2012]{Little}
Little, S.~E. (2012).
\newblock {\em Female Reproduction}, chapter~40, pages 1195--1227.
\newblock Published Online.

\bibitem[Loss et~al., 2013]{Loss}
Loss, S.~R., Will, T., and Marra, P.~P. (2013).
\newblock The impact of free-ranging domestic cats on wildlife of the united
  states.
\newblock {\em Nature communications}, 4:1396.

\bibitem[Loyd et~al., 2013a]{Loyd}
Loyd, K.~A., Hernandez, S., Abernathy, K., Shock, B., and Marshall, G. (2013a).
\newblock Risk behaviours exhibited by free-roaming cats in a suburban us town.
\newblock {\em The Veterinary record}, 173.

\bibitem[Loyd et~al., 2013b]{Loyd2}
Loyd, K.~A., Hernandez, S., Carroll, J., Abernathy, K., and Marshall, G.
  (2013b).
\newblock Quantifying free-roaming domestic cat predation using animal-borne
  video cameras.
\newblock {\em Biological Conservation}, 160:183–189.

\bibitem[McCarthy, 2019]{Mccarthy2}
McCarthy, R. (2019).
\newblock Traditional surgical and laparoscopic vasectomy in dogs and cats.
\newblock {\em Clinical Theriogenology}, 11(3):243--246.

\bibitem[McCarthy et~al., 2013]{Mccarthy}
McCarthy, R., Levine, S., and Reed, J. (2013).
\newblock Estimation of effectiveness of three methods of feral cat population
  control by use of a simulation model.
\newblock {\em Journal of the American Veterinary Medical Association},
  243:502--11.

\bibitem[McCune, 2010]{McCune}
McCune, S. (2010).
\newblock {\em The Domestic Cat}, chapter~31, pages 453--472.
\newblock John Wiley \& Sons, Ltd.

\bibitem[Mella-Méndez et~al., 2022]{Mella}
Mella-Méndez, I., Flores~Peredo, R., Amaya-Espinel, J., Bolívar-Cimé, B.,
  Mac Swiney~González, M.~C., and Martínez, A. (2022).
\newblock Predation of wildlife by domestic cats in a neotropical city: a
  multi-factor issue.
\newblock {\em Biological Invasions}, 24.

\bibitem[Mendes-de Almeida et~al., 2006]{Mendes}
Mendes-de Almeida, F., Faria, M., Landau-Remy, G., Branco, A., Barata, P.,
  Chame, M., Pereira, M., and Labarthe, N. (2006).
\newblock The impact of hysterectomy in an urban colony of domestic cats (felis
  catus linnaeus, 1758).
\newblock {\em Int J Appl Res Vet Med}, 4:134--141.

\bibitem[Miller et~al., 2014]{Miller2}
Miller, P., Boone, J., Briggs, J., Lawler, D., Levy, J., Nutter, F., Slater,
  M., and Zawistowski, S. (2014).
\newblock Simulating free-roaming cat population management options in open
  demographic environments.
\newblock {\em PloS one}, 9:e113553.

\bibitem[Monraz~Ibarra, 2023]{MonrazAbandono2023}
Monraz~Ibarra, M.~A. (2023).
\newblock Proposición con punto de acuerdo que exhorta a la secretaría de
  hacienda y crédito público para que, en coordinación con la secretaría de
  medio ambiente y recursos naturales, concreten una estrategia dentro del
  paquete económico para el ejercicio fiscal 2024, a fin de que consideren:
  crear un programa presupuestario que garantice el apoyo a refugios para
  animales domésticos y realizar deducciones sobre el impuesto sobre la renta
  de productos de alimento para mascotas y servicios médicos veterinarios.
\newblock Gaceta del Senado.

\bibitem[Munson and Moresco, 2007]{Munson}
Munson, L. and Moresco, A. (2007).
\newblock Comparative pathology of mammary gland cancers in domestic and wild
  animals.
\newblock {\em Breast disease}, 28:7--21.

\bibitem[Nutter et~al., 2004]{Nutter}
Nutter, F., Levine, J., and Stoskopf, M. (2004).
\newblock Reproductive capacity of free-roaming domestic cats and kitten
  survival rate.
\newblock {\em Journal of the American Veterinary Medical Association},
  225:1399--402.

\bibitem[Ogan and Jurek, 1997]{Ogan}
Ogan, C. and Jurek, R.~M. (1997).
\newblock Biology and ecology of feral, free-roaming, and stray cats.
\newblock Pages 87-92 in J.E. Harris, and C.V. Ogan, (eds.), Mesocarnivores of
  northern California: biology, management, and survey techniques, workshop
  manual. August 12-15, 1997, Humboldt State University, Arcata, CA. The
  Wildlife Society, California North Coast Chapter, Arcata, CA 127 p.

\bibitem[Overley et~al., 2004]{Overley}
Overley, B., Shofer, F., Goldschmidt, M., Sherer, D., and Sorenmo, K. (2004).
\newblock Association between ovarihysterectomy and feline mammary carcinoma.
\newblock {\em Journal of veterinary internal medicine / American College of
  Veterinary Internal Medicine}, 19:560--3.

\bibitem[Patrón~Laviada, 2022]{PatronAbandono2022}
Patrón~Laviada, C.~A. (2022).
\newblock Proposición con punto de acuerdo por el que se exhorta al titular
  del ejecutivo federal a través de la secretarfa de medio ambiente y recursos
  naturales y de la procuraduría federal de protección al ambiente, para que
  dentro de sus facultades elabore un plan de acción a nivel nacional para la
  esterilización de animales de compañía, a fin de beneficiar a los animales
  en situación de calle y controlar la sobrepoblación de animales
  abandonados.
\newblock Gaceta del Senado.

\bibitem[Perusquía, 2021]{Perus}
Perusquía, R. (2021).
\newblock Regulación de una población en estado feral desde la perspectiva de
  la teoría de control.
\newblock Tesina. Posgrado en Ciencias Matemáticas, UNAM.

\bibitem[Ramos-Rendón et~al., 2023]{Ramos}
Ramos-Rendón, A., Gual-Sill, F., Cervantes, F., González-Salazar, C.,
  García-Morales, R., and Martínez-Meyer, E. (2023).
\newblock Assessing the impact of free-ranging cats (felis silvestris catus)
  and dogs (canis lupus familiaris) on wildlife in a natural urban reserve in
  mexico city.
\newblock {\em Urban Ecosystems}, 26:1--14.

\bibitem[Robertson, 2008]{Robertson}
Robertson, S.~A. (2008).
\newblock A review of feral cat control.
\newblock {\em Journal of Feline Medicine and Surgery}, 10(4):366--375.

\bibitem[Rodriguez~Ferrere, 2022]{Rodriguez}
Rodriguez~Ferrere, M.~B. (2022).
\newblock Animal welfare underenforcement as a rule of law problem.
\newblock {\em Animals : an open access journal from MDPI}, 12(11):1411.

\bibitem[Rodríguez-Rodríguez et~al., 2022]{RodriguezRodriguez}
Rodríguez-Rodríguez, E.~J., Gil-Morión, J., and Negro, J.~J. (2022).
\newblock Feral animal populations: Separating threats from opportunities.
\newblock {\em Land}, 11(8):1370.

\bibitem[Satz, 2009]{Satz}
Satz, A. (2009).
\newblock Animals as vulnerable subjects: Beyond interest-convergence,
  hierarchy, and property.
\newblock {\em Derecho Animal. Forum of Animal Law Studies}, 1.

\bibitem[Sharpe, 2016]{Sharpe}
Sharpe, J. (2016).
\newblock {\em A Mathematical Model for Feral Cat Ecology With Application To
  Disease}.
\newblock Electronic Thesis and Dissertations 2004-2019.

\bibitem[Stornelli, 2007]{Stornelli}
Stornelli, M. (2007).
\newblock Particularides fisológicas de la reproducción en felinos.
\newblock {\em Revista Brasilera de Reproducción Animal}, 31:71--76.

\bibitem[Thompson et~al., 2022]{Thompson}
Thompson, B.~K., Sims, C., Fisher, T., Brock, S., Dai, Y., and Lenhart, S.
  (2022).
\newblock A discrete-time bioeconomic model of free-roaming cat management: A
  case study in knox county, tennessee.
\newblock {\em Ecological Economics}, 201:107583.

\bibitem[Tsutsui et~al., 2009]{Tsutsui}
Tsutsui, T., Higuchi, C., Soeta, M., Oba, H., Mizutani, T., and Hori, T.
  (2009).
\newblock Plasma lh, ovulation and conception rates in cats mated once or three
  times on different days of oestrus.
\newblock {\em Reproduction in domestic animals = Zuchthygiene}, 44 Suppl
  2:76--8.

\bibitem[Veronesi and Fusi, 2022]{Veronesi}
Veronesi, M. and Fusi, J. (2022).
\newblock Feline neonatology: From birth to commencement of weaning – what to
  know for successful management.
\newblock {\em Journal of Feline Medicine and Surgery}, 24:232--242.

\bibitem[Villagrán~Villasana, 2021]{VillagranAbandono2021}
Villagrán~Villasana, A.~J. (2021).
\newblock Iniciativa con proyecto de decreto por el que se adicionan un
  artículo al código penal para el distrito federal, en materia de abandono
  animal.
\newblock Congreso Nacional de la Ciudad de México, II Legislatura.

\end{thebibliography}
\end{document}